\documentclass[12pt]{article}
\usepackage{amsmath}
\usepackage{latexsym}
\usepackage{amssymb}
\usepackage{a4wide}
\usepackage{bbm}
\usepackage{epsfig}
\newcommand{\I}{\text{i}}

\newcommand{\Tr}{\text{Tr}}

\newcommand{\re}[1]{~(\ref{#1})}

\newcommand{\Gk}{\Gamma_k}
\newcommand{\Gt}{\Gamma_k^{(2)}}

\newcommand{\cbar}{\bar{C}}
\newcommand{\Lam}{\Lambda}
\newcommand{\LUV}{\Lambda_{\text{UV}}}
\newcommand{\pat}{\partial_t}
\newcommand{\gqb}{g_\Lambda^2}
\newcommand{\gqh}{\hat{g}_\Lambda^2} 
\newcommand{\mqk}{{m}_k^2}
\newcommand{\Nc}{N_{\text{c}}}

\newcommand{\PT}{P_{\text{T}}{}}
\newcommand{\PL}{P_{\text{L}}{}}

\newcommand{\gb}{g_\Lambda}
\newcommand{\gh}{\hat{g}_{\Lambda}}
\newcommand{\xh}{\hat{x}}
\newcommand{\yh}{\hat{y}}
\newcommand{\zh}{\hat{z}}

\newcommand{\ZT}{Z}
\newcommand{\za}{\mathcal{Z}_A}
\newcommand{\zc}{\mathcal{Z}_C}

\newcommand{\vDt}{\tilde{v}_D}
\newcommand{\Kreg}{K_{\text{reg}}}

\newcommand{\cV}{\mathcal{V}}

\newcommand{\beqa}{\begin{eqnarray}}
\newcommand{\eeqa}{\end{eqnarray}}
\newcommand{\beq}{\begin{equation}}
\newcommand{\eeq}{\end{equation}}

\begin{document}
$\text{}$

\vspace{-2.3cm}

{\hfill \small\sf HD-THEP-04-29, {  }IPPP/04/17, {  }DCPT/04/34, 
{  }http://arXiv.org/abs/hep-ph/0408089} 

\vspace{1.5cm}

\centerline{\Large\bf Renormalization flow of Yang-Mills propagators}



\vspace{.8cm}



\centerline{Christian S.~Fischer${}^1$ and Holger Gies${}^2$}

\vspace{.6cm}
\centerline{\small\it${}^1$ Institute for Particle Physics Phenomenology,
University of Durham,}
\centerline{\small\it South Rd, Durham DH1 3LE, UK}
\centerline{\small\it  E-mail:  christian.fischer@durham.ac.uk}
\centerline{\small\it${}^2$ Institut f\"ur theoretische Physik,
  Universit\"at Heidelberg,}
\centerline{\small\it Philosophenweg 16, D-69120 Heidelberg,
  Germany}
\centerline{\small\it  E-mail:  h.gies@thphys.uni-heidelberg.de}

\begin{abstract}
We study Landau-gauge Yang-Mills theory by means of a nonperturbative
vertex expansion of the quantum effective action. Using an exact
renormalization group equation, we compute the fully dressed gluon and
ghost propagators to lowest nontrivial order in the vertex
expansion. In the mid-momentum regime, $p^2\sim\mathcal{O}(1)\text{GeV}^2$, 
we probe the propagator flow with various {\em ans\"atze} for the three- 
and four-point correlations. We analyze the potential of these truncation 
schemes to generate a nonperturbative scale. We find universal infrared 
behavior of the propagators, if the gluon dressing function has developed 
a mass-like structure at mid-momentum. The resulting power laws in the 
infrared support the Kugo-Ojima confinement scenario.
\end{abstract}

\section{Introduction}

A remarkable property of four-dimensional Yang-Mills theories is the
nonperturbative generation of a scale, separating the physics at high
energies from that of low energies. Many high-energy phenomena can
well be controlled with the elaborate machinery of perturbation
theory. By contrast, essential low-energy properties such as
confinement and bound-state formation as well as the transition region
between the nonperturbative and the perturbative regime, are not yet
fully understood.

\noindent
\quad Genuinely nonperturbative frameworks such as functional methods for
computing Green's functions or lattice Monte-Carlo simulations are
required to investigate these phenomena. As a main advantage,
functional continuum methods can cover orders of magnitude in momentum
space and therefore naturally connect the high- and low-momentum
regime. Analytical solutions are available in both the ultraviolet
(UV) and also the infrared (IR) region of momentum. Certainly, the
drawback of functional approaches to the infinite set of Green's
functions is that one has to  
rely on truncations to obtain a closed, solvable system of equations.

A prominent representative of the functional approach is the tower of
Dyson-Schwinger equations (DSE) that has found a variety of
applications in the context of QCD 
\cite{Alkofer:2000wg,Roberts:2000aa,Maris:2003vk}. 
In Landau-gauge Yang-Mills theory, the lowest order n-point functions,
the ghost and gluon propagators, have been investigated in various 
approximation and truncation schemes in the DSE approach 
\cite{Alkofer:2000wg,vonSmekal:1997is,Atkinson:1998tu,Lerche:2002ep,
Zwanziger:2001kw,Fischer:2002hn,Fischer:2003rp}. The picture emerging
in the IR is  
that the 'geometric' degree of freedom, i.e., the Faddeev-Popov 
determinant, dominates over the dynamics of the gluon field: the ghost
and gluon propagators are described by simple power laws with an IR 
finite or even vanishing gluon propagator (depending on the truncation 
scheme) and a ghost propagator more singular than a simple pole. In the 
momentum region accessible to lattice simulations, such a picture is nicely 
confirmed \cite{Suman:1995zg,Cucchieri:1997fy,Langfeld:2001cz,
Gattnar:2004bf,Bonnet:2001uh,Bowman:2004jm}.

These IR power laws are in agreement with both Zwanziger's horizon 
condition arising from the gauge-fixing procedure and the Kugo-Ojima 
confinement criterion. The horizon condition is formulated as a boundary 
condition on the ghost and gluon propagators. It restricts the integration 
of gauge-field configurations in the Schwinger functional to the 
Gribov region, defined by a positive semidefinite Faddeev-Popov operator
\cite{Zwanziger:1993qr}. It has been shown that this restriction is 
sufficient to ensure that physical expectation values are not affected by 
gauge copies \cite{Zwanziger:2003cf}. Entropy arguments have been employed 
to reason that the IR modes of the gauge and ghost fields are close 
to the Gribov horizon. The resulting horizon conditions state that the 
gluon propagator should vanish in the IR and the ghost propagator 
should be more singular than a simple pole \cite{Zwanziger:2002ia}.  

The Kugo-Ojima confinement criterion \cite{Kugo:1979gm,Nakanishi:qm} is a 
dynamical condition on the two-point function 
$\langle D_\mu c D_\nu \bar{c} \rangle$, which ensures the conservation of 
global color charge. Provided BRST symmetry holds also nonperturbatively, 
one can then define a physical state space $\cV_{phys}$ containing 
colorless states only. In Landau gauge the Kugo--Ojima criterion can be 
translated into a condition for the IR behavior of the ghost 
propagator: it is fulfilled if the ghost propagator is more singular than 
a simple pole \cite{Kugo:1995km}. What remains to be shown in this 
scenario is the appearance of a mass gap and the violation of cluster 
decomposition in $\cV_{phys}$.

Functional methods for computing Green's functions can be implemented in
various different but interrelated formulations, each with its own
assets and drawbacks. In this work, we employ the exact
renormalization group (RG) \cite{Wegner,Wetterich:yh,Reuter} formulated in
terms of flow equations for the n-point functions. The exact RG flow 
is derived from the scaling behavior of the effective action with 
respect to an IR cutoff $k$. Starting with an initial condition for 
the effective action at a UV cutoff scale $\LUV$, e.g., 
in terms of the bare action, the RG flow equation is integrated 
from $k=\LUV$ down to $k=0$, thereby generating the full quantum 
effective action. If the effective action is expanded in powers 
of fields, these couplings are the (1PI) Green's functions 
(proper vertices) of the theory. 

The RG equations are, for instance, intimately related to the 
DSEs.\footnote{There is also a close
relation to the functional approach based on 2PI (or, more generally,
$n$PI) effective actions which are currently investigated in
the context of non-equilibrium quantum field theory \cite{2PI}.}
On a formal level, it can be shown 
that effective actions satisfying the DSEs in the presence of an IR 
cutoff $k$ are (quasi-)fixed points of the exact RG flow. Thus a 
solution of the flow equation for $k \rightarrow 0$ is also a solution 
of the ordinary DSEs \cite{Ellwanger:1996wy,Terao:2000ae}. This property 
should hold in an approximate sense even if the effective action is 
truncated, provided the truncation preserves the dominant structure of 
the theory. 

In comparison to the corresponding DSEs, the flow equation for the Yang-Mills 
propagators has been much less explored. The first investigation was
reported in \cite{Ellwanger:1995qf,Ellwanger:1996wy}. From the behavior of the
propagator functions around $\Lambda_{\text{QCD}}$, a strong IR
divergence of the gluon propagator consistent with the old idea of
infrared slavery was conjectured. Similar observations were made in
\cite{Bergerhoff:1997cv}.  These results seem to be in marked contrast
to the IR power laws described above; however, in these works, the
integration of the flow was only performed down to a finite value of
$k$, and the deep IR $k\to0$ was not explored. Indeed, a recently
developed flow-equation technique for performing a 
self-consistency study of the IR scaling of propagators
finds power laws in agreement with the ones obtained in the
DSE formulation, thereby reconciling the two approaches
\cite{Pawlowski:2003hq}. 
 
In this work, we employ the RG flow equation in order to study the
full flow from the perturbative UV to the deep IR, investigating the
possible mechanisms that connect the high-energy behavior of the
propagators with their IR power laws. In particular, we will not use
self-consistency criteria for determining the IR power laws, but solve
the flow in order to see if and how the power laws emerge.  The paper
is organized as follows: in the following, we first summarize aspects
of Landau-gauge Yang-Mills theory that are relevant to our work. In
section \ref{flow} we introduce the exact RG equations for the
propagators of Yang-Mills theory and discuss our initial truncation of
the effective action. The role of gauge symmetry in controlling the
flow is specified.  In section \ref{results}, we first discuss
important qualitative features of the flow equation and then present
our numerical results. We focus in particular on the delicate
mid-momentum regime, $p^2\sim\mathcal{O}(1)\text{GeV}^2$, by probing
the propagator flow with various vertex {\em ans\"atze}. Finally, we
solve the flow towards the IR, in order to explore the ``domain of
attractivity'' for which the IR power laws represent an IR stable
fixed point. This again provides information about the dynamical
mechanisms required in the mid-momentum regime, which still remains
insufficiently understood.  We end with a detailed discussion of our
results in the conclusions.

\section{Propagators in Landau-gauge Yang-Mills theory}\label{LG-YM}

In the continuum Green's function approach to Yang-Mills theory, Landau gauge has been
a favorite choice for a number of reasons. First, of all linear covariant gauges, Landau 
gauge is the only one symmetric under the exchange of ghosts and antighosts. This, on the
one hand, allows one to interpret ghosts and antighosts as (unphysical) particles and 
antiparticles. On the other hand, this symmetry is a useful guiding principle for the
construction of a nonperturbative ansatz for the ghost-gluon vertex 
\cite{Lerche:2002ep,Alkofer:2003jr}. 
As will be detailed later on, such an ansatz is necessary in both the DSE and the exact RG
flow-equation framework in order to close the associated equations determining the ghost 
and gluon propagators. Secondly, the ghost-gluon vertex does not
suffer from UV  
divergencies in Landau gauge \cite{Taylor:ff}, i.e., the vertex renormalization factor
$Z_{\cbar A C}$ can be set to one. Again, this provides a useful constraint 
on possible nonperturbative vertex dressings. In fact, as has been shown in the 
DSE framework \cite{Atkinson:1998tu,Lerche:2002ep,Zwanziger:2001kw,Fischer:2002hn}, 
even the bare ghost-gluon vertex is a good approximation in the UV and IR.

Thirdly, a direct consequence of UV finiteness of the ghost-gluon vertex in 
Landau gauge is a nonperturbative definition of the running coupling in terms of the propagators.
In Euclidean momentum space, the ghost and gluon propagators are given by
\beqa
D_{\bar{C}C}(p^2) &=& - \frac{G(p^2)}{p^2}, \\
D_{AA}(p^2)       &=& \left(\delta_{\mu \nu} -\frac{p_\mu p_\nu}{p^2}\right) \frac{Z(p^2)}{p^2},
\eeqa
with the ghost dressing function $G(p^2)$ and the gluon dressing
$Z(p^2)$.\footnote{Here we employ the conventions of the DSE
literature; in the RG literature, the dressing functions are usually
inversely defined: e.g., $Z(p^2)\to 1/Z(p^2)$.} With the help of the STI,
\beq
1 = Z_{\bar{C}AC} = Z_g\: Z_C\:Z_A^{1/2}, \label{norm-sti}
\eeq
relating the vertex renormalization factor with the corresponding factors for the coupling 
$g$, the ghost and the gluon fields, we can define a running coupling $\alpha(p^2)$ by
\cite{Mandelstam:1979xd,vonSmekal:1997is}
\beq
\alpha(p^2) = \frac{g^2}{4\pi}\:G^2(p^2)\:Z(p^2). \label{coupling}
\eeq
Note that the right-hand side of this definition is an RG invariant, i.e., 
$\alpha(p^2)$ does not depend on the renormalization scale. Within the Green's functions 
approach, the IR behavior of the ghost and gluon dressing functions can be determined 
analytically. For momenta $p \ll \Lambda_{QCD}$, the dressing functions are given by 
simple power laws,
\beq
G(p^2) \sim (p^2)^{-\kappa}, \quad \quad Z(p^2) \sim (p^2)^{2 \kappa}, \label{power}
\eeq
with interrelated exponents \cite{vonSmekal:1997is,Atkinson:1998tu,Lerche:2002ep,
Zwanziger:2001kw,Fischer:2002hn,Pawlowski:2003hq}. 
Hereby $\kappa$ is an irrational number, $\kappa = (93 - \sqrt{1201})/98 \approx 0.595$, 
which depends slightly on the truncation scheme \cite{Lerche:2002ep}. From the power laws,
Eq.~(\ref{power}), we see immediately that the running coupling, Eq.~(\ref{coupling}), 
has a fixed point in the IR.

A fourth reason why Landau gauge is interesting has already been mentioned in the 
introduction: in this gauge, there is a direct connection between the Kugo-Ojima confinement 
scenario and the ghost propagator. If the ghost propagator is more singular than a simple 
pole, global gauge symmetry is unbroken and one can demonstrate that the physical state 
space of the theory, defined as the cohomology of the BRST operator, contains color 
singlets only \cite{Kugo:1995km}.

Finally, Landau gauge is known to be a fixed point of the
renormalization flow \cite{Litim:1998qi}. Within a given truncation, it
is possible to show that this fixed point is IR attractive for a wide
range of initial gauge parameters \cite{Ellwanger:1995qf}. This
suggests that an investigation of Yang-Mills theory starting in a
general linear covariant gauge at large IR cutoff will
generically end up in the Landau gauge after the flow has been
integrated. This, together with the simplifications mentioned above,
speaks for Landau gauge as a natural and convenient choice from the
very beginning.

\section{Flow equation for the vertex expansion}
\label{flow}

\subsection{Exact renormalization group}

To provide a short summary of the exact RG approach,
let us begin with the gauge-fixed action for SU($\Nc$) Yang-Mills
theory in a covariant gauge in $D$ dimensional Euclidean spacetime,
\begin{equation}
S_{\text{YM}}=\int d^D x\, \left[ \frac{1}{4} F_{\mu\nu}^a
F_{\mu\nu}^a + \frac{1}{2\xi} (\partial_\mu A_\mu^a)^2 + (\partial_\mu
\cbar^a)  D_\mu^{ab}(A) C^b \right].\label{2.1}
\end{equation}
In the flow equation approach, an IR cutoff $k$ for the quantum
fluctuations is implemented by adding a Gau\ss ian term to the action,
\begin{equation}
\Delta S_k=\frac{1}{2} \int d^D x\, A_\mu^a R_{A, \mu\nu}^{ab} A_\nu^a 
	+ \int d^D x\, \cbar^a R_{C}^{ab} C^b.\label{2.2}
\end{equation}
The momentum-dependent regulators $R_{A,C}$ cut off the IR
fluctuations for gluons and ghosts, respectively, at a momentum scale
$p^2\simeq k^2$. Defining the quantum theory via the Schwinger
functional with the action $S_{\text{YM}}+\Delta S_k$, a Legendre
transform leads us to the effective average action $\Gk$. It already 
contains the effects of all quantum fluctuations with momenta larger
than $k$, and governs the dynamics of the remaining modes with
momenta smaller than $k$. The response of the effective action under
a variation of the cutoff scale $k$ is described by the flow equation,
\begin{equation}
\pat \Gk[\phi]=\frac{1}{2}\, \Tr\, \mathcal{G}_A[\phi] \pat R_A 
	-\Tr\, \mathcal{G}_C[\phi] \pat R_C, \label{2.3}
\end{equation}
where $\phi=(A,C,\cbar)$ and $t=\ln k/\Lam$. The traces run over all
indices including momentum, and
$\mathcal{G}_{A/C}[\phi]=(\Gt[\phi]+R)^{-1}_{A/C}$ abbreviates the full
regularized gauge/ghost field propagator with $\Gt[\phi]=\delta^2
\Gk[\phi]/\delta \phi\delta\phi$. Once initial conditions are
specified at a high scale $\Lambda$ in terms of the microscopic action
to be quantized, $\Gamma_{k=\Lambda}\to S_{\text{YM, bare}}$, the flow
equation describes the RG trajectory of the effective average action
towards the IR. The end point $\Gamma=\Gamma_{k=0}$
corresponds to the full quantum effective action. The flow equation is
UV and IR finite by construction, it is exact despite its
one-loop structure, and the momentum trace is dominated by momenta $p^2\simeq
k^2$. 

For a good approximate solution of the flow equation, the ansatz for
the effective action should include the relevant degrees of freedom
for the problem under consideration. In this work, we assume that an
expansion of the effective action in terms of full vertices can
illuminate aspects of the nonperturbative structure of the theory,
\begin{equation}
\Gk[\phi]= \sum_{n}\frac{1}{n\!}\, \int_{p_1,\dots,p_n}\,
\delta(p_1+\dots+p_n) \, \Gamma_k^{(n)} (p_1,\dots,p_n)\, \phi(p_1) \dots
\phi(p_n), \label{2.4}
\end{equation}
where $\int_p =\int d^D p/(2\pi)^{D}$, and $\delta(p)=(2\pi)^{D}
\delta^{(D)}(p)$. Inserting Eq.\re{2.4} into Eq.\re{2.3} and taking
appropriate derivatives, we obtain an
infinite set of coupled first-oder differential equations for the
proper vertices $\Gamma_k^{(n)}$. Truncating the expansion at order
$n_{\text{max}}$ leaves all equations for the vertices
$\Gamma_k^{(n\leq n_{\text{max}}-2)}$ unaffected. In order to close
this tower of equations, the vertices of order $n_{\text{max}}$ and
$n_{\text{max}}-1$ can either be derived from their truncated
equations or taken as bare -- or even built upon inspired
guesswork. This defines a consistent approximation scheme that can in
principle be iterated to arbitrarily high orders in $n_{\text{max}}$.

\subsection{Truncated flow equations}

In this work, we set $n_{\text{max}}=4$ and solve the dynamic equations
for the 2-point vertices: the inverse propagators. To this end, we 
employ either bare 3- and 4-point vertices or use constructions derived
from further physical information. The tree-level-vertex structure of 
the Landau gauge is exploited, e.g., by setting the four-ghost vertex 
to zero. Our truncation for the effective action is given by
\begin{eqnarray}
\Gk&=&\frac{1}{2}\int_q A_\mu^a(-q) \left[ \frac{q^2}{\ZT(q^2)}\,
  \PT_{\mu\nu} + \frac{q^2}{Z_{\text{L}}(q^2)}\, \PL_{\mu\nu}
	+ \mqk \delta_{\mu\nu} \right]  A_\nu^a(q)
	+\int_q \cbar^q(q)\, \frac{q^2}{G(q^2)}\, C^a(q), \nonumber\\
&&+\I \gb\int_{q_1,q_2,q_3} \delta(q_1+q_2+q_3)\,  V_{\cbar
  AC,\, \mu}^{abc}(q_1,q_2,q_3)\, \cbar^a(-q_1) A_\mu^b(q_2) C^c(q_3)
      \nonumber\\
&&+\I\gb\int_{q_1,q_2,q_3} \delta(q_1+q_2+q_3)\,
  V_{3A,\, \mu}^{abc}(q_1,q_2,q_3)\, A_\nu^a(q_1) A_\mu^b(q_2)
  A_\nu^c(q_3)   \nonumber\\
&& +\frac{1}{4} \gb^2 \int_{q_1,\dots q_4} \delta(q_1+\dots+q_4)
  V_{4A}^{abcd} (q_1,\dots,q_4) A_\mu^a(q_1)A_\nu^b(q_2)
  A_\mu^c(q_3)A_\nu^a(q_4), \label{2.5}
\end{eqnarray}
where $\gb$ denotes the renormalized gauge coupling at the high scale
$\Lam$ where our flow will be initiated. In the first line, the
$k$-dependent dressing functions $\ZT(p^2),Z_{\text{L}}(p^2),G(p^2)$
characterize the fluctuation-induced modifications of the bare
propagators of the transversal and longitudinal gluons and the ghosts,
respectively. These are at the center of interest in the present
work. Furthermore, a gluon mass $\mqk$ has been included explicitly,
the role of which will be explained in detail later on. 

\begin{figure}[h]
\centerline{
\epsfig{file=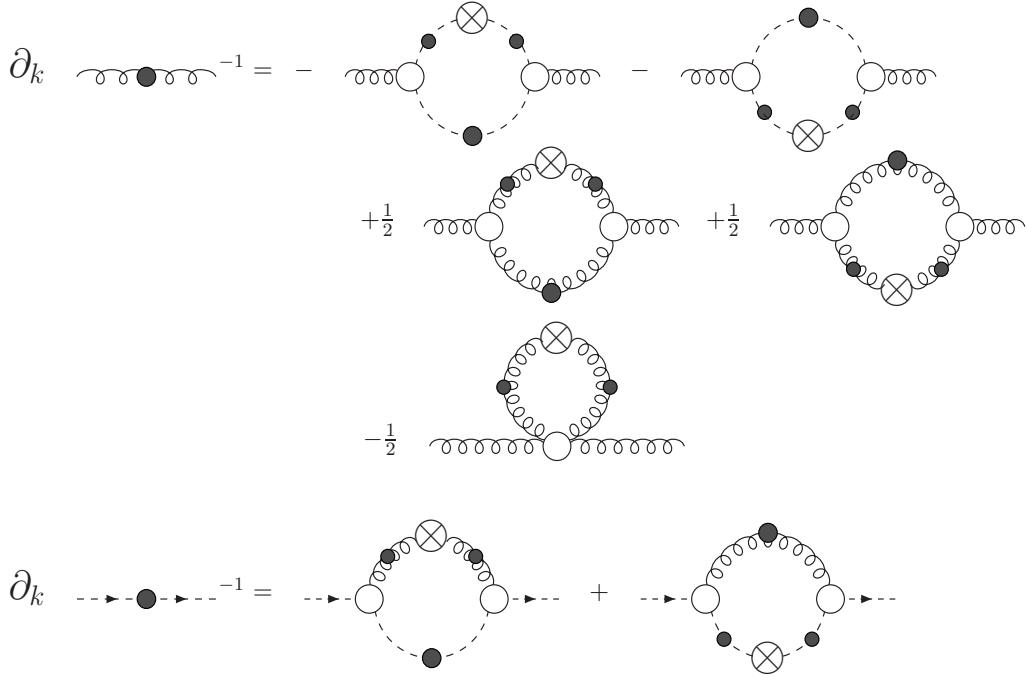,width=14cm,height=9cm}
}
\caption{Flow equations for the propagators. Filled circles denote
full propagators, open circles correspond to full vertices, and
crossed circles indicate the $\pat R_k$-operator insertion.} \label{diag}
\end{figure}

Inserting this truncation into Eq.\re{2.3}, the flow equation reduces
to a set of equations for the propagators which we display
diagrammatically in Fig.~\ref{diag}.
In the following, we will consider truncations where the Lorentz and
color index structures of the vertices are kept identical to the bare
ones:
\beqa
V_{\cbar AC,\mu}^{abc}(q_1,q_2,q_3) &=& f^{abc} q_{1\mu}
	V_{\cbar AC}(q_1,q_2,q_3) \nonumber\\ 
V_{3A,\mu}^{abc}(q_1,q_2,q_3)&=&f^{abc} q_{1\mu}\, V_{3A}(q_1,q_2,q_3) \nonumber\\ 
V_{4A}^{abcd}(q_1,\dots,q_4)&=&f^{eab}f^{ecd} \, V_{4A}(q_1,\dots,q_4). \label{2.6}
\eeqa
Here $V_{\cbar AC}, V_{3A}, V_{4A}$ are functions of the momentum
invariants of $q_{i\mu}$, examples of which will be given later; the
bare-vertex truncation is defined by $V_{\cbar AC}=V_{3A}=V_{4A}=1$ up
to RG rescalings of the fields (cf. Appendix \ref{scaling}). 

Even though exact solutions of the flow equation are independent of
the choice of the regulator, approximate solutions do generally
depend on the specific form of $R$. A careful choice of the
regulator improves numerical stability as well as the quality of 
the truncation in general \cite{Litim:2001up}. 
Here we concentrate on regulators of the type
\begin{equation} 
R_A(p^2)=\left(\frac{p^2}{\ZT(p^2)}\,
\PT(p)+\frac{p^2}{Z_{\text{L}}(p^2)}\, \PL(p) \right)\,  r_A(p^2/k^2),
\quad
R_C(p^2)=\frac{p^2}{G(p^2)}\, r_C(p^2/k^2). \label{2.7}
\end{equation}
The dimensionless regulator shape function $r_{A,C}(\xh)$ with a 
dimensionless argument determines the analytic form of the
momentum space regularization. In addition, we have included the
dressing functions in the regulator. In the first place, this
construction guarantees that the flow equations are invariant under 
RG rescaling \cite{Pawlowski:2001df}. 
Moreover, since we expect a fast RG running of many
couplings in the nonperturbative domain, the inclusion of the dressing
functions accounts for an adjustment of the regulator to the spectral
flow of quantum fluctuations \cite{Ellwanger:1996wy,Litim:2002xm,Gies:2002af}.

Let us now introduce some notation in order to facilitate a more
explicit representation of the flow equation for the propagators.
First we specify squared momentum variables, $x,y,z$, involving the external
momentum $p$ and the loop momentum $q$,
\begin{eqnarray}
&&x=p^2,\quad y=q^2,\quad z=(p-q)^2=x+y-2u\sqrt{xy}, \quad
(\xh,\yh,\zh)=(x,y,z)/k^2, \label{xyz.def}
\end{eqnarray}
where $u:=p.q/(|p||q|)$ denotes an angle variable. Here and in the
following all hatted quantities are 
dimensionless by appropriate rescaling with $k^2$. 
Although we are interested in $D=4$ dimensional spacetime in this
work, it is worthwhile for future application to perform the
computation in arbitrary $D$ dimensional spacetime. For this, we need
the $D$ dimensional loop integral of an arbitrary function of $x,y,u$:
\begin{eqnarray}
\int \frac{d^Dq}{(2\pi)^D}\, f(x,y,u) &=&\vDt\, \int_0^\infty dy\,
y^{\frac{D}{2} -1} \int_{-1}^1 du\, (1-u^2)^{\frac{D-3}{2}}\, f(x,y,u),
\label{Dint}\\
&&\vDt=\frac{2}{\sqrt{\pi}}\,
\frac{\Gamma(\frac{D}{2})}{\Gamma(\frac{D}{2} -\frac{1}{2})}\, v_D,
\quad v_D=\frac{1}{2^{D+1} \pi^{D/2}\, \Gamma(D/2)}. \label{vD.def}
\end{eqnarray}
Here $v_D$ is related to the surface volume of a $D$ dimensional
sphere $S^{D-1}$. In order to display all dimensionful quantities in
terms of $k^2$, it is also useful to introduce the dimensionless
coupling $\gh$,
\begin{equation}
\gqh=\gqb\, k^{D-4}, \label{gh.def}
\end{equation}
where $\gb$ denotes the value of the coupling at an initial high scale
$\Lam$.

After insertion of this specific regulator form given above, the flow
equation for the inverse transverse gluon propagator
$\Gamma^{(2)}_{k,AA}(p^2)=p^2/\ZT(p^2)+\mqk$ can be written as 
\begin{eqnarray}
\pat\left( \frac{x}{\ZT(x)}+\mqk\!\right)
&=&-4 \gqh \Nc \vDt k^2 \int_0^\infty d\yh\, \yh^{\frac{D}{2} -1}
	\int_{-1}^1 du\, (1-u^2)^{\frac{D-3}{2}} 
 \nonumber\\
&&\qquad \times \bigg[ Q^{\ZT,\ZT}(\xh,\yh,u) \,
\Kreg^{\ZT,\ZT} (\xh,\yh,u)\,V_{3A}\,\ZT(\yh k^2) \ZT(\zh k^2)\, V_{3A}
 \nonumber\\
&&\qquad\qquad\quad +   Q^{G,G}(\xh,\yh,u) \,
   \Kreg^{G,G} (\xh,\yh,u)\, V_{\cbar AC}\, G(\yh k^2) G(\zh k^2)
  V_{\cbar AC}\bigg]\nonumber
\\
&&-4\frac{(D-1)^2}{D} \gqh\Nc v_D\, k^2 \int_0^\infty d\yh\,
\yh^{\frac{D}{2}-1} \label{fm1} \\
&&\qquad\qquad \times\frac{\ZT(\yh k^2)}{p_A(\yh)} \, 
\bigg(-\yh^2 r_A'(\yh) -\frac{1}{2} \yh r_A(\yh) \pat \ln \ZT(\yh
k^2)\bigg)\, V_{4A},
\nonumber
\end{eqnarray}
where the last term is the tadpole contribution. Since we are ultimatly
aiming at the Landau gauge, we have omitted any contribution of the
longitudinal modes on the right-hand side of the flow equation which
decouple in this limit. The
flow equation for the inverse ghost propagator $\Gamma^{(2)}_{k,\bar C
C}(p^2)=p^2/G(p^2)$ then yields
\begin{eqnarray}
\pat \frac{x}{G(x)}&=&2 \gqh \Nc \vDt k^2 \!\int_0^\infty\!\!
d\yh\, \yh^{\frac{D}{2} -1} \!\int_{-1}^1\! \!du\,
(1-u^2)^{\frac{D-3}{2}} 
\bigg[  Q^{G,\ZT}(\xh,\yh,u) \,
\Kreg^{G,\ZT} (\xh,\yh,u) \nonumber\\
&&\qquad\qquad\qquad\qquad\qquad\qquad\qquad
 \,\times V_{\cbar AC}\, G(\yh k^2) \ZT(\zh k^2)\, 
V_{\cbar AC}
 \!\bigg]\!.
\label{fm2} 
\end{eqnarray}
The quantities $Q^{\ZT,\ZT}$, $Q^{G,G}$ denote the kernels of the 
gluon and ghost loop, respectively, in the gluon flow equation, whereas
$Q^{G,\ZT}$ represents the ghost equation kernel. These kernels are 
given in Appendix \ref{kernels}.

The quantity $\Kreg$ abbreviates the contribution from the regularized
propagators which reads:
\begin{eqnarray}
\Kreg^{Z_a,Z_b}(\xh,\yh,u)&=&\frac{1}{p_a(\yh) p_b(\zh)} 
\bigg[\frac{-\yh^2 r_a'(\yh) -\frac{1}{2} \yh r_a(\yh) \pat\ln Z_a(\yh
k^2)}{p_a(\yh)} \nonumber\\
&&\qquad\qquad\qquad +
\frac{-\zh^2 r_b'(\zh) -\frac{1}{2} \zh r_b(\zh) \pat\ln Z_b(\zh
k^2)}{p_b(\zh)}\bigg], \label{Kreg}
\end{eqnarray}
where the prime denotes a derivative with respect to the argument.
Here $Z_{a,b}$ denote gluon and ghost dressing functions, $\ZT(p^2)$ and $G(p^2)$, 
and $p_{a,b}$ abbreviate the corresponding regularized inverse propagators,
\begin{equation}
p_C(\yh)=\yh \: (1+r_C(\yh)),\quad 
p_A(\yh)=\yh \: (1+r_A(\yh))+ \ZT(\yh k^2) \: \mqk/k^2. \label{invprop}
\end{equation}
An explicit expression for the regulator shape function can be 
found in Appendix \ref{cutoffs}.

Note that all scale derivatives $\pat\equiv k (d/dk)$ that occur are
taken at fixed dimensionful {\em external} momentum, $\pat x = 0=\pat
(\xh k^2)$, so that the scale derivatives on both sides of the flow
equation have the same meaning. However, on {\em internal} momentum
variables the scale derivative acts on the manifest $k^2$ dependence,
$\pat \yh=0$. This implies the somewhat clumsy differentiation rule
for the dependent momentum variable $\pat (\zh k^2)=2\yh k^2 -2u
\sqrt{\yh k^2 x}$.

Equations\re{fm1} and\re{fm2} represent a closed set of equations for
the gluon and ghost propagators in the Landau gauge. Once we have
specified initial conditions for $Z$ and $G$ at, say, a
perturbative UV scale $k=\Lambda$, we can integrate the flow equations
down to $k=0$ and read off the form of the fully dressed
propagators. However, before we do so, we first have to discuss the
issue of gauge invariance. 

\subsection{Gauge invariance \label{wtisec}}

Since we are working in a gauge-fixed formulation of Yang-Mills
theory, gauge invariance of the system is encoded in a constraint for
the effective action. This constraint can either be formulated as a
Ward-Takahashi identity or, invoking the BRST formalism, as a
Slavnov-Taylor identity. In addition to the breaking of gauge
invariance by the gauge-fixing procedure, the regulator term\re{2.2}
represents another source of gauge or BRST symmetry breaking. In order
to account for this additional breaking, both the Slavnov-Taylor and
the Ward-Takahashi identity are modified by further terms depending on
the regulator. These guarantee the restoration of gauge invariance in
the limit $k \rightarrow 0 $.\footnote{As an alternative to
the present formalism, the construction of manifestly gauge-invariant
flows has been put forward in \cite{Morris:1999px}.} The BRST formalism 
in the flow equation framework has been worked out in 
Refs.~\cite{Reuter,Ellwanger:iz,Bonini:1994kp}. Since the resulting equation
is important for the results of the present paper, we rederive the
same findings from the Ward-Takahashi identity here, following
\cite{Freire:2000bq}.

Using the momentum-space representation of the generator
$\mathcal{G}^a$ of infinitesimal gauge transformations formulated in
terms of unrenormalized coupling $g_0$ and the fields
$A_{0\mu}^a,C^a_0,\cbar^a_0$ which occur in the microscopic action $S$,
\begin{equation}
\mathcal{G}^a(p)=\I p_\mu \frac{\delta}{\delta A_{0\mu}^a(-p)} -g_0
f^{abc}  \int_q \left[A_{0\mu}^c(q) \frac{\delta}{\delta
	A_{0\mu}^b(q-p)}  +C_0^c(q)\frac{\delta}{\delta C_0^b(q-p)}
+\cbar_0^{c}\frac{\delta}{\delta \cbar_0^b (q+p)} \right], \label{Ga.def}
\end{equation}
the modified Ward-Takahashi identity (mWTI) for covariant gauges can
be written as\footnote{Here we use a convenient representation of the
mWTI in terms of connected n-point functions to be evaluated from the
Schwinger functional in presence of the regulator term\re{2.2}; of
course, by Legendre transformation, the connected n-point functions
can be expressed in terms of the effective action $\Gamma_k[\phi]$ and
derivatives thereof. Moreover, our notation does not distinguish
between the fluctuation fields (to be integrated over in the Schwinger
functional) and the corresponding ``classical'' field (conjugate to
the source), but the meaning should be obvious from the context. }
\begin{eqnarray}
&&\mathcal{G}^a(p)\left[ \Gamma_k -\int \left(\frac{1}{2\xi}
	(\partial_\mu A_{0,\mu}^a)^2 + (\partial_\mu \cbar_0^a)  
	D_\mu^{ab}(A_0) C_0^b \right) \right] 
\nonumber\\
&&\quad =-\frac{1}{\xi} g_0 f^{abc} \int_q (p+q)_\mu (p+q)_\nu 
\langle A_{0,\mu}^c(-q) A_{0,\nu}^b(p+q) \rangle_{\text{con}} \nonumber\\
&&\qquad-g_0 f^{abc} \int_{q_1,q_2} p_\mu \big( \delta(p+q_1-q_2)
q_{2\mu} \delta^{bf} -\I g_0 f^{bef}A_{0,\mu}^e(p+q_1-q_2)\big) \langle
\cbar_0^c(q_1) C_0^f(q_2)\rangle_{\text{con}} \nonumber\\
&&\qquad -\I g_0^2 f^{abc}f^{feb} \int_{q_1,q_2} p_\mu\langle
\cbar_0^c(q_1) A_{0,\mu}^e(p+q_1-q_2) C_0^f(q_2)\rangle_{\text{con}}
\nonumber\\ 
&&\qquad
-\frac{1}{2} g_0 f^{abc}\int_q\big[R_{A,\mu\nu}(p+q)-R_{A,\mu\nu}(q)\big] 
	\langle A_{0,\mu}^c(-q) A_{0,\nu}^b(p+q) \rangle_{\text{con}}
\nonumber\\
&&\qquad
-g_0 f^{abc}\int_q\big[R_{C}(p+q)-R_{C}(q)\big] 
	\langle\cbar_0^c(q) C_0^b(q+p)\rangle_{\text{con}}, \label{mWTI}
\end{eqnarray}
where the first four lines represent the standard Ward-Takahashi
identity, and the last two lines are the modification owing to the
regulator. In the limit $k\to0$, the regulator terms vanish such that
the standard WTI is recovered.  The important observation is that the
mWTI is compatible with the flow equation in the sense that if a
solution of the flow equation satisfies the mWTI at one scale, it does
so for all scales. Hence, if our initial conditions at the high scale satisfy
the mWTI at $k=\Lam$, the end point of the RG trajectory at $k\to 0$
which is the full quantum effective action will be gauge invariant,
i.e., will fulfill the standard WTI. However, these statements only
hold for the full effective action. A truncated flow can leave the
gauge-invariant trajectory which is constrained by the mWTI. In order
to satisfy the symmetry principle encoded in the mWTI, the degrees of
freedom in the truncation can be subdivided into truly {\em
independent} ones, the dynamics of which is determined by the flow
equation, and the {\em dependent} ones that can be expressed in terms
of independent degrees of freedom by virtue of the mWTI. In this way,
the symmetry principle is satisfied on the theory subspace defined by
the truncation, and gauge invariance is consistently implemented
within the truncation \cite{Ellwanger:iz,Litim:1998nf}.

In the present truncation, the mWTI first of all constrains the
longitudinal gluon propagator. Although the longitudinal components
decouple in the Landau gauge which we are aiming at, the mWTI
nevertheless provides an important piece of information. Evaluating
Eq.\re{mWTI} in the zero-momentum limit of the longitudinal
propagator, the mWTI constrains the value of the gluon mass $m_k$. In order
to represent this constraint for the gluon mass in terms of the
quantities appearing in our truncation\re{2.5}, we need to know the
relation between the fields $(A_{0},\cbar_0,C_0)$ appearing in
$\mathcal{G}^a$ and the fields renormalized at the scale $\Lam$ in which
the effective action is expressed, i.e., we need to determine the 
renormalization factors $Z_A$ and $Z_C$ defined by
\begin{equation}
A_{0\mu}=\sqrt{Z_A} A_\mu,\quad 
(\cbar_0,C_0)=\sqrt{Z_C}\,(\cbar,C). \label{Zfactors}
\end{equation}
The connection between the renormalization factors in the exact RG approach 
and the conventional $\overline{\text{MS}}$ scheme have been worked 
out in Ref. \cite{Ellwanger:1997tp}. Identifying $\Lam$ with
the renormalization scale, the boundary conditions for the flow
at $\Lam$ are equivalent to renormalization conditions in the 
$\overline{\text{MS}}$ scheme, and the finite parts of the respective 
renormalization factors can be related to each other.  

As will be detailed in Sect.~\ref{PertIC}, our initial effective action 
at $\Lam$ will be obtained by integrating the bare action 
$S_{\text{YM}}+\Delta S_k$, Eqs.(\ref{2.1},\ref{2.2}), from an even larger scale 
$\LUV$ down to $\Lam$. With 
\begin{equation}
1=Z_{T,\LUV}(0) = Z_{\Lam}(0) Z_A(\Lam),\quad
1=G_{\LUV}(0) = G_{\Lam}(0) Z_C(\Lam),
\end{equation}
we then obtain
\begin{equation}
A_\mu=\sqrt{Z_{\Lambda}(0)} A_{0\mu},\quad 
(\cbar,C)=\sqrt{G_\Lambda(0)}\,(\cbar_0,C_0), \label{Gzero}
\end{equation}
where $(Z_{\Lambda}(p^2),G_\Lambda(p^2))=\lim_{k\to\Lambda}
(\ZT(p^2),G(p^2))$. Furthermore, the nonperturbative definition of the
running coupling, Eq.\re{coupling}, gives the relation
\begin{equation}
g_0^2=\gqb\, G_\Lambda^2(0) Z_{\Lambda}(0). \label{gbare}
\end{equation}
With these relations the mWTI for the gluon mass yields \cite{Ellwanger:1996wy}
\begin{eqnarray}
m_k^2\!\!\!&=&\!\!\!\frac{8}{D} v_D \gqh\Nc\, G_\Lambda(0)\,  k^2 \int_0^1 d\yh\, 
	\yh^{D/2 -1} 
   \Bigg\{V_{\cbar AC}\, G(\yh k^2) \frac{(-\yh^3
	r_C'(\yh))}{p_C^3(\yh)}  +(D\!-\!1)\frac{V_{3A}\,\ZT(\yh
	k^2)}{p_A^2(\yh)}\nonumber\\ 
&&\!\!\!\times	\left[ \frac{D\!-\!1}{4} \yh r_A(\yh) + \frac{\yh^3
	r_A'(\yh)}{p_A(\yh)} + \frac{Z(\yh k^2) m_k^2}{k^2} \frac{(\yh^2
	r_A'(\yh)+\yh r_A(\yh))}{p_A(\yh)} +\frac{Z'(\yh k^2) m_k^2}{k^2}
	\frac{\yh^2 r_A(\yh)}{p_A(\yh)}\!\right]\!\!\!\Bigg\}\!,
	\nonumber\\ 
&&\label{mWTIm}
\end{eqnarray}
where we have again dropped longitudinal contributions on the
right-hand side according to the Landau gauge, and primes denote a
derivative with respect to $\yh$.  The occurrence of the gluon mass 
$m_k$ is an obvious consequence of gauge-symmetry breaking by the regulator. 
Assuming that the dressing functions diverge, if at all, with a simple
power or weaker $V_{3A}Z(p^2),V_{\cbar AC}G(p^2)\lesssim
(1/p^2)^\kappa$ in the IR limit, the $k$ dependence of the mass is
bounded by $m_k^2\lesssim (k^2)^{1-\kappa}$ towards the IR.  Removing
the regulator with $k\to 0$, the mass vanishes as long as
$\kappa<1$. Therefore, $\kappa<1$ serves as a consistency condition
that our solutions have to obey in order to satisfy the standard WTI
in the present truncation.

Since the gluon mass also appears on the right-hand side
of Eq.\re{mWTIm}, the mWTI represents a ``gap'' equation for a
self-consistent determination of the gluon mass. Moreover, the RG
evolution of this mass is completely determined by the evolution of
$\ZT$ and $G$ and the vertices. In this way, the mWTI guarantees that
the gluon mass is not an RG relevant parameter, as is
conventionally the case for boson masses, but RG irrelevant, since it
drops to zero with $k\to 0$.

On the other hand, from the flow equation for the transversal gluon
propagator\re{fm1}, we can read off the flow of the gluon mass in the
zero-momentum limit. Of course, for the exact solution, this mass flow
would fully agree with that of the mWTI; both mass equations also
agree in the perturbative limit of the truncation ($G,\ZT\to\,$const.,
$V_{\cbar AC,3A,4A}\to 1$), which serves as a nontrivial check of the
approach \cite{Ellwanger:1996wy}. Nonperturbatively, however, both equations for the
mass can differ strongly, even qualitatively, since an exact
implementation of the gauge constraint always requires information
which is beyond the truncation. The gluon mass is particularly
sensitive to this fact, since any deviation from a gauge-invariant RG
trajectory can lift its protection against quadratic renormalization
(as can occur for massive bosons); then, the gluon mass would be of
the order of the cutoff, $m_k^2\sim \Lam^2$. Indeed, this is what would
happen if we took the mass flow of the transversal gluon seriously,
disregarding the mWTI. In similar truncations of DSEs, the very same
fact occurs as quadratic divergencies in the gluon equation which have
to be removed by hand. In the context of flow equations, we have the
mWTI at our disposal for controlling the mass flow. For this, we
subtract the mass flow from the transversal gluon equation and put
in the mass from the mWTI:
\begin{equation}
\pat \frac{x}{\ZT(x)}=\text{RHS\re{fm1}}\big|_{m_k^2\,(\text{mWTI})}
 	- \pat	 m_k^2\big|_{\text{flow}} 
  =\text{RHS\re{fm1}}\big|_{m_k^2\,(\text{mWTI})} 
	- \text{RHS\re{fm1}}|_{x\to0}, \label{masscor}
\end{equation}
where the gluon mass in the RHS is taken from the mWTI. For the exact
solution, Eq.\re{masscor} corresponds to a zero operation, whereas in
the truncation, this procedure removes quadratically renormalizing
parameters by virtue of the gauge constraint.\footnote{In
\cite{Ellwanger:1996wy}, the difference of the propagator flow
resulting from using either the mWTI mass flow or the direct mass flow
was taken as a dynamical criterion for the breakdown of the
truncation. In view of the qualitatively different RG behavior of the
two masses in this truncation, this criterion is rather restrictive.
The flow mass corresponds to an unphysical quadratic divergence in
this truncation and hence can be expected to deviate significantly
from the mWTI mass. Here we abandon this criterion and give preference
to the mWTI mass.}

\section{Flow analysis}\label{results}

In this section, we analyze the flow equation partly analytically but
mainly numerically for various initial conditions and truncations. We
concentrate on the case of $D=4$ dimensional spacetime and $\Nc=3$
colors unless stated otherwise. In the next subsection we will specify 
two different types of boundary conditions at large and intermediate 
momenta that serve as initial conditions for the flow. We proceed 
by analyzing the RG equations analytically on a qualitative level and determine
conditions for the higher n-point functions such that the IR attractive 
power-law domain can be approached by the flow. Guided by these findings,
we then perform a numerical analysis of the flow in various truncation 
schemes. The details of our numerical methods are given in Appendix 
\ref{numproc}.

\subsection{Initial conditions}
\label{PertIC}

The solution of a quantum theory is determined by the RG flow equation
and the corresponding initial conditions. For the latter, we have to
specify the details of the initial effective action $\Gamma_\Lam$,
serving as our boundary condition at the scale $\Lam$. For a perturbatively
small coupling $g_\Lambda$ at $\Lam$, one could choose the bare action
(\ref{2.1}) for $\Gamma_\Lam$. Improved initial conditions can be
obtained by perturbatively integrating the flow equation from a larger
scale $\LUV$ down to $\Lam$, employing bare dressing functions on the
right-hand side of the flow equation. This guarantees the correct
one-loop behavior of the dressing functions and subsequently of the
running coupling for momenta $p^2 \gg \Lam$ far larger than our
starting scale. Of course, a correct matching to the full flow at
$\Lam$ requires the inclusion of the regulator function $R_k$ in this
perturbative determination of the initial conditions. Qualitatively,
the high-momentum behavior, $p^2\gg\Lam^2$, of the dressing functions
for one-loop improved perturbative initial conditions (PIC) is 
\begin{equation}
\ZT^{\text{PIC}}(p^2)|_\Lam\simeq \ZT_\Lam \left( 1+\gamma_A\,
\frac{11}{\Nc \alpha_\Lam}{12 \pi} \ln \frac{p^2}{\Lam^2}\right), \quad
G^{\text{PIC}}(p^2)|_\Lam\simeq G_\Lam \left( 1+\gamma_C\, \frac{11}{\Nc
\alpha_\Lam}{12 \pi} \ln \frac{p^2}{\Lam^2} \right), 
\label{IC1}
\end{equation}
where $\gamma_A=-13/22$ and $\gamma_C=-9/44$ denote the anomalous
dimensions for gluons and ghosts, respectively, for arbitrary $\Nc$,
and $\ZT_\Lam,G_\Lam=\ZT(\Lam^2),G(\Lam^2)|_\Lam$ denote the
normalizations of the fields. For low momenta $p^2\ll\Lam^2$, the
perturbative initial conditions for the dressing functions go to a
constant, owing to the presence of the regulator.\footnote{In
principle, we could also use improved initial conditions in the form
of one-loop leading-log resummed perturbation theory. However, 
such an improvement would hardly exert any influence on the 
remaining flow towards the IR, which is the focus of the
present investigation.}

Certainly, universality tells us that such a one-loop RG improvement
is irrelevant for the questions related to the IR behavior.  Whether a
truncation is good enough to stabilize the flow in the mid-momentum
regime between the perturbative UV and the deep IR
should not depend on the details of the UV behavior of the
initial effective action. We have confirmed that this is indeed not
the case, employing both a bare initial action and its one-loop
improved counterpart.

Since it turns out that a full solution connecting perturbative
initial conditions with the expected IR behavior is difficult to
construct, we design instead a set of mid-scale initial conditions for
exploring in particular the flow in the nonperturbative IR regime. For
this, we consider the initial scale $\Lam$ to be already in the
nonperturbative regime but still high enough such that the IR
asymptotics has not yet developed, say, for instance, $\Lam\gtrsim
1$GeV. For definiteness, we study two contrary cases: one in which the
ghost dressing function has already developed a scale $L$, whereas the
gluon remains perturbative (or even constant),
\begin{equation}
G^{(\eta_C)}(p^2)|_\Lam=G_\Lam\,
\left(\frac{p^2}{L^2+p^2}\right)^{\eta_C}, \quad 
\ZT^{(\eta_C)}(p^2)|_\Lam=\ZT_\Lam. \label{IC2}
\end{equation}
Here, $\eta_C$ denotes a trial exponent to be varied, and the scale $L$ 
separates perturbative from deep IR behavior. The other case is given 
by a scale in the gluon dressing function and a perturbative ghost,
\begin{equation}
G^{(\eta_A)}(p^2)|_\Lam=G_\Lam, \quad
\ZT^{(\eta_A)}(p^2)|_\Lam=\ZT_\Lam\,
\left(\frac{p^2}{L^2+p^2}\right)^{\eta_A}. \label{IC3} 
\end{equation}
For a trial gluon power $\eta_A \sim 1$, the scale $L$ plays the role 
of a nonperturbative gluon mass. Such a nonperturbative gluon 
mass should not be confused with the mass $m_k$ controlled by the WTI, 
cf. subsection \ref{wtisec}. In general, the scale $L$ will survive
when the IR cutoff $k$ is taken to zero, whereas $m_k \rightarrow 0$ 
in this limit according to the WTI. 

Note that Eqs.\re{IC2} and\re{IC3} are initial conditions at a fixed
mid-momentum scale, and do not represent an {\em ansatz} for the form
of the dressings at even lower scales. In other words, we will
determine the resulting full momentum dependence at lower scales and
not merely the flow of the parameters $\eta_{A,C}$ and $L$.

The use of these mid-scale initial conditions \re{IC2},\re{IC3} 
for the IR flow facilitates 
a study of the IR attractive domain of the asymptotic
power solutions obtained so far in the literature. In this way, we can
analyze the necessary ingredients for a solution of the flow at an
intermediate scale $k$ to finally run into the power-law solutions for
$k\to 0$; it dispenses us from the need to solve the complete flow
from the UV to the IR in particular in the transition region, where
vertex corrections are expected to become important.

\subsection{Qualitative analysis}\label{cond-flow}

Some important features of the present set of RG equations for the
propagators can already be extracted by a qualitative analysis. We
begin with the observation that the perturbative initial conditions
\re{IC1} are such that the dressing functions are increasing functions
for decreasing momentum $p^2=x$. In view of the expected IR behavior
$G(x)\sim x^{-\kappa}$ with $\kappa\gtrsim0.5$ (cf. Sect.~\ref{LG-YM}), 
this property should
persist and even be enhanced for the ghost dressing. By contrast, the
gluon dressing is expected to decrease in the IR with $\ZT(x)\sim
x^{2\kappa}$, such that the flow equation has to bend the gluon
dressing in the IR.  

Let us check the ghost equation\re{fm2} first. As long as the vertex
dressing $V_{\cbar AC}$ remains positive, the RHS of the ghost
flow is strictly positive; particularly, the kernel $Q^{G,G}$, 
given in Appendix \ref{kernels}, and the dressing  functions are positive. 
The first term in the regulator quantity $\Kreg$ is positive as
$r^\prime(\hat{y}) <0$ for all monotonous regulator shape functions,
like the one specified in Eq.~(\ref{2.8}) below. Moreover, the mechanism 
for a sign change should not depend on the details of the regulator. 
The other terms $\sim \pat \ln Z_a$ in $\Kreg$  are positive for 
perturbative initial conditions, and thus they cannot induce a sign
change, owing to the $\pat$ structure. Regarding the RHS of  
the ghost equation as ``$\beta$
functions'' for a set of ``couplings'' $x/G(x)$ labeled by $x$, the
positivity of these ``beta functions'' guarantees that the
``couplings'' decrease towards the IR. Consequently, this decrease of
$x/G(x)$ with $k\to 0$ implies an increase of $G(x)$ towards the
IR. This is consistent with the IR expectation of an enhanced ghost,
$G(x)\sim x^{-\kappa}$.

Next we apply this argument to the gluon equation\re{fm1}. Since we
anticipate a suppression of $\ZT(x)$ for small $x$, we need 
a negative ``$\beta$ function'' for $x/\ZT(x)$. Hence, we expect a
negative RHS for small $x$, once we have subtracted the gluon mass
flow $\pat m_k^2$ in order to isolate the flow of $x/\ZT(x)$. Let us
analyze Eq.\re{fm1} term by term, beginning with the gluon loop with
kernel $Q^{\ZT,\ZT}$. For simplicity, we consider first
the limit of $x\ll k^2$ where the propagator and vertex dressings as
well as $\Kreg$ become rather independent of $x$, owing to the
regularization (the $\hat{z}$ dependence can safely be replaced by
$\yh$ in these quantities). Hence the sign of the kernel after mass
flow subtraction determines the overall sign. Using
representation\re{A.1}, this corresponds to
\begin{equation}
Q^{\ZT,\ZT}(\xh,\yh,u)
  -Q^{\ZT,\ZT}(0,\yh,u)=-(1-u^2)\left[ \xh+
\frac{\xh\yh}{\zh}\left(1-\frac{1-u^2}{D-1}\right) \right]\leq 0.
\label{qa1}
\end{equation}
Together with the overall minus sign in Eq.\re{fm1}, we find that the
gluon loop contribution to the ``$\beta$ function'' is positive. Hence
$x/\ZT(x)|_{\text{gluon loop}}$ decreases towards the IR, implying
that $\ZT(x)|_{\text{gluon loop}}$ increases. We have derived this
result in the limit $x\ll k^2$; but owing to the regularization, a
well-converging expansion of the flow in $x/k^2$ exists, such that the
validity of this statement can be extended to $x\lesssim k^2$. 
Even within the limits of these mild assumptions, we arrive at an important
first result: the gluon loop in the gluon equation cannot be the
source of a bending and a subsequent suppression of the gluon
dressing. This statement holds for all 3-gluon vertex dressings that
are non-negative and preserve the bare-vertex index structure.

Now we turn to the ghost-loop contribution to the gluon equation with
kernel $Q^{G,\ZT}$. The situation here is more subtle,
since the kernel is independent of $x$ (see Eq.\re{A.2}), such that
the residual terms after mass-flow subtraction arise from the
$\zh$-dependent ghost and vertex dressings and $\Kreg$. Collecting all
these $\zh$ dependencies in a function $f(\zh)\sim V_{\cbar A C}
G(\zh k^2) \Kreg$, we have to study the sign of ($\zh|_{x\to0}=\yh$):
\begin{equation}
f(\zh)-f(\yh) = \xh f'(\yh) + \mathcal{O}( \xh^2), \label{qa2}
\end{equation}
where we have dropped all terms odd in $u$ that are canceled by the
$u$ integration. $G(p^2)$ as well as $\Kreg$ for generic regulator
shape functions are decreasing functions of the momentum. In the
bare-vertex truncation $V_{\cbar A C}=1$, the RHS of Eq.\re{qa2} is
therefore negative for small $\xh$, because $f'(\yh)\leq 0$. Together
with the overall minus sign of Eq.\re{fm1}, this again implies a
positive ``$\beta$ function'' and thus a contribution to the flow of
$\ZT(x)$ that cannot suppress the dressing function in the IR. This
statement is, of course, weaker than the one given above for the gluon
loop, since a non-negative ghost-gluon vertex dressing can, in
principle, turn the sign of $f'(\yh)$. However, this would be an
unexpected source for the bending of the gluon dressing, since the
ghost loop is not anticipated to play a dominant role at intermediate
scales. 

Let us finally continue our discussion with the tadpole contribution
in Eq.\re{fm1}. For a bare-vertex truncation $V_{4A}\to$ const., the
tadpole does not contribute at all to the flow of $\ZT(x)$, but only to
the mass flow. However, as soon as the 4-vertex acquires a nontrivial
momentum dependence under the flow, the tadpole can contribute with
either sign. Especially if the vertex dressing decreases with  
external momentum $x$, its ``$\beta$ function'' contribution can be
negative, thereby inducing a suppression of the gluon in the IR. 

To summarize, a vertex expansion of the effective action has the
potential to describe Landau-gauge ghost enhancement and gluon
suppression in agreement with, e.g., lattice results. A realization of
this scenario in a truncation on the 3-vertex level requires
exceptional (i.e., negative) vertex dressings or suitable index
structures beyond the bare ones. By contrast, a truncation including
dressed 4-vertices can accommodate gluon suppression more
conventionally with merely positive vertex dressings and bare index
structures.\footnote{Yet another option by which the observations of
this section could be evaded is given by the freedom of performing
$k$-dependent field redefinitions under the flow
\cite{Gies:2001nw}. This property of the flow can be exploited in
order to optimize the degrees of freedom in a truncation. However,
in the absence of a convincing optimization criterion, we will not explore this
possibility any further in the present work.}

\subsection{Vertices}
\label{vertices}

Since we intend to perform a vertex expansion to lowest non-trivial
order, we do not attempt to solve the full dynamical equations for the
vertices. The choice of the latter hence determines the remaining
unspecified part of the truncation. Basically three different
strategies can be pursued: first, the highest vertices in a vertex 
expansion can be taken as bare. Second, these vertices can be
computed self-consistently from their truncated dynamical equations. 
These first two strategies define a consistent approximation scheme 
generalizable to higher orders. Third, the vertices can be modeled 
employing further physical information.

In the literature, the second strategy of deriving or at least
constraining vertices by their truncated equations has been frequently
followed. In particular in gauge theories, the constraints from the
Ward-Takahashi or Slavnov-Taylor identities can serve as an additional
input and have been exploited in
\cite{Ellwanger:1996wy,Ellwanger:1995qf,vonSmekal:1997is}.
However, it still remains unclear whether this strategy can be
successful, since the truncated equations may lack important
structural information from the neglected terms; in fact, this manner of
construction can even lead to inconsistencies as exemplified in
\cite{Atkinson:1998tu,Lerche:2002ep,Watson:1999ha}.

In this work, we use a set of different vertices in order to explore
different routes that the flow can take from the UV to the IR. First
of all, the ghost-gluon vertex is taken to be bare, $V_{\cbar A C}=1$,
which is in agreement with the non-renormalization
property\re{norm-sti} but neglects possible finite
renormalizations. In the gluonic sector, we consider: 
\begin{itemize}

\item[(a)] bare 3-gluon vertex, $V_{3A}=G_\Lam/\ZT_\Lam$, where
factors $G_\Lam,\ZT_\Lam$ take the normalization of the fields at
$\Lam$ into account (cf. Appendix \ref{scaling});

\item[(b)] modified bare 3-gluon vertex, $V_{3A}=\rho \,
G_\Lam/\ZT_\Lam$, where the possibly $k$-dependent parameter $\rho$
can be varied in 
order to suppress or enhance the 3-gluon vertex; we will especially
make use of a vertex suppression with $\rho$ becoming gradually
smaller than 1, in agreement with recent lattice results
\cite{Boucaud:2003xi}. An exceptional case is given by $\rho=0$:
the ``ghost-loop-only'' truncation;

\item[(c)] dressed 3-gluon vertex, 
\begin{equation}
V_{AAA}=\frac{1}{G(q^2)^{1+3\delta} \, Z(q^2)^{(1+3\delta)/2}}\,
\frac{1}{G((p-q)^2)^{1+3\delta} \, Z((p-q)^2)^{(1+3\delta)/2}}
,\label{vertex_c}  
\end{equation}
where $q^2$ and $(p-q)^2$ denote the momenta running around the loop,
and $\delta=-9/44$ is the anomalous dimension of the ghost dressing
function to one-loop order. This type of vertex construction has been
used in the DSEs to ensure the correct one-loop
anomalous dimensions of the ghost and gluon dressing functions in the
UV \cite{Fischer:2002hn}. Since it breaks RG scaling
invariance explicitly in the nonperturbative momentum regime, 
rescaling the initial conditions for the fields
with $G_\Lam,\ZT_\Lam$  corresponds to suppressing or enhancing this
vertex analogous to (b);

\item[(d)] bare 4-gluon vertex $V_{4A}=G_{\Lam}^2/\ZT_\Lam$. However, since the
4-gluon vertex occurs only in the tadpole diagram, the bare-4-gluon
vertex truncation is identical to the no-4-gluon-vertex truncation, since
the momentum-independent part of the tadpole contributes only to the
mass flow and thus is subtracted completely;

\item[(e)] dressed 4-gluon vertex, 
\begin{equation}
V_{4A}= G^{\gamma_1}(p^2) G^{\gamma_2}(r^2)\, \ZT^{\zeta_1}(p^2)
\ZT^{\zeta_2}(r^2), \quad \gamma_1+\gamma_2=2, \quad
\zeta_1+\zeta_2=-1, \label{v4A}
\end{equation}
where $p^2$ denotes the external momentum, and $r^2$ can be set equal
to the internal momentum $q^2$ or other scales such as $k^2$ or
$\Lam^2$. The constraints for the exponents $\gamma_i$,
$\zeta_i$ arise from RG rescaling invariance. 

\end{itemize}

\subsection{Numerical results}

\subsubsection{Perturbative initial conditions and bare vertices}

Here we solve the flow from the perturbative UV
towards the IR, imposing one-loop improved initial
conditions. For illustration, we start the flow at $\Lam=M_Z=91.187$
GeV with $\alpha(M_Z)=0.118$.

\begin{figure}
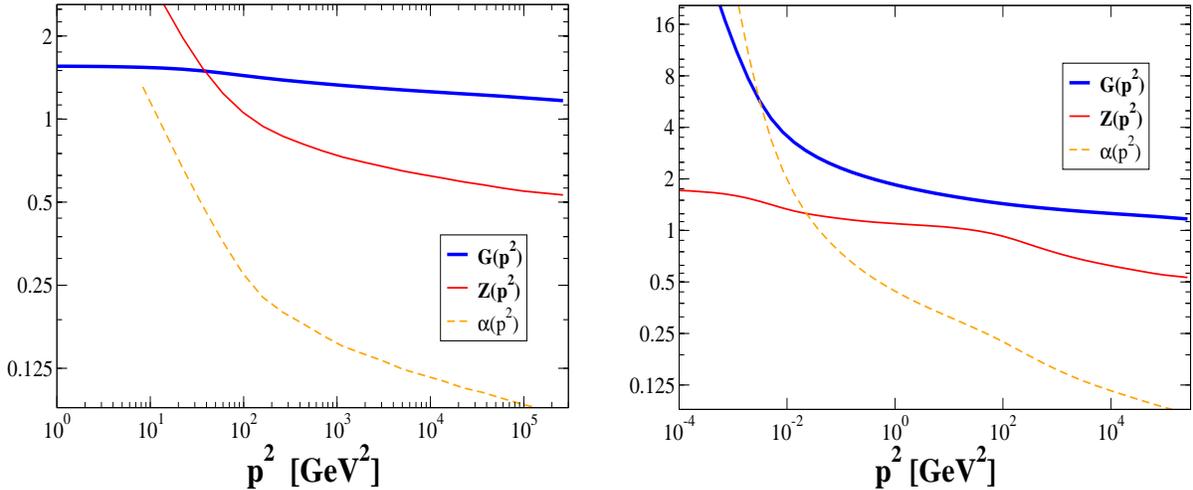

\centerline{
\epsfig{file=pert.run.bare.eps,width=7.5cm,height=6.5cm}
\hspace{0.5cm}
\epsfig{file=pert.run.mbare.eps,width=7.5cm,height=6.5cm}}
\caption{ Numerical results for ghost and gluon dressing functions and
the running coupling employing the bare vertex truncation. Left panel:
bare 3-gluon vertex (a). Right panel: modified bare 3-gluon vertex (b)
with $\rho$ decreased from 1 to 0.2. Both flows become singular at a
finite value of $k$, signaling the breakdown of the oversimplified truncation. The
coupling exhibits a singularity of Landau-pole type. Other
multiplicative positive 3-gluon vertex dressings lead to similar
results.}
\label{bare-v}
\end{figure}

Our results for the bare-vertex truncation are shown in
Fig.~\ref{bare-v}. In agreement with our analysis of
Sect.~\ref{cond-flow}, both dressing functions increase towards the
IR. After a few orders of magnitude of perturbative running, the flow
in fact becomes singular at a finite scale $k_\text{sing}$ where the
gluon dressing diverges.\footnote{This gluon divergence should not be
confused with a $1/p^4$ behavior of the gluon propagator, as
conjectured earlier in the context of infrared slavery. The
divergence of the flow at $k_{\text{sing}}$ implies that the gluon
propagator diverges at a finite value of $p^2$. Hence this divergence
signals a breakdown of the truncation and not the onset of a $1/p^4$
behavior.} As a consequence, the coupling $\alpha$ also diverges
reminiscent of the Landau-pole singularity of perturbation theory.

This singular behavior of the gluon is mainly driven by the
gluon loop involving the 3-gluon vertex. Since the latter is expected
to be suppressed in the IR, let us use the modified bare 3-gluon
vertex with a suppression parameter of $\rho<1$. On the left panel of
Fig.~\ref{bare-v}, we show the flow for this truncation with $\rho$
gradually decreased from 1 to 0.2 for decreasing $k$. Though the gluon
dressing now remains finite, it is the ghost dressing that diverges at
finite $k$ and thereby triggers a singularity of Landau-pole type. 

These observations are in agreement with the analysis of
Sect.~\ref{cond-flow} and stress the fact that a truncation with a
multiplicative positive dressing of the 3-gluon vertex is not capable
of describing the flow from the perturbative UV to the IR power-law
asymptotics. 
 
\subsubsection{Perturbative initial conditions with dressed 4-gluon
vertex}
\label{dress4}

Let us now include a momentum-dependent 4-vertex dressing of the type
(e) in Eq.\re{v4A}. In agreement with our observations of
Sect. \ref{cond-flow}, the truncation now has the potential to
describe an IR suppression of the gluon towards the expected IR
asymptotics. 

\begin{figure}
\centerline{
\epsfig{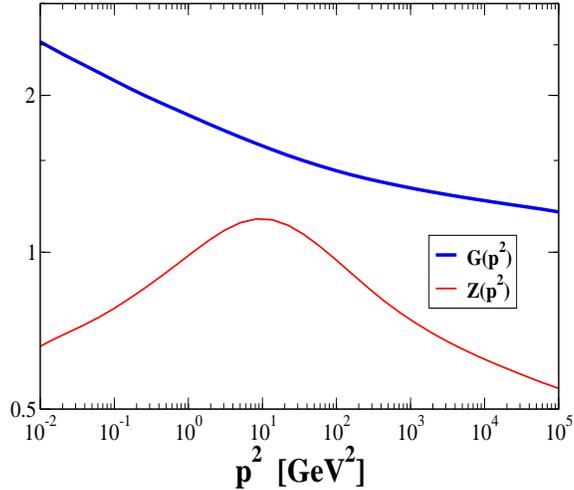}
}
\caption{ Numerical results for the ghost and gluon dressing functions,
including a dressed 4-gluon vertex (e). This truncation is capable of
describing ghost enhancement and gluon suppression in the IR. (Vertex
parameters: $\gamma_1=1.1$, $\zeta_1=1/2$, $r^2=\Lam^2$, and
$\rho=1\dots 1/2$.)}
\label{figv4A}
\end{figure}

In Fig. \ref{figv4A}, we display a typical solution for the gluon and
ghost dressings, exhibiting IR ghost enhancement and gluon
suppression. All solutions look qualitatively similar for vertex
parameters $\gamma_1,\zeta_1>0$ in Eq.\re{v4A}. We also observe a
quantitative stability for $\gamma_1=\mathcal{O}(1)$ and $\zeta_1=0.1
\dots 0.5$. (Fig. \ref{figv4A} is computed with $\gamma_1=1.1$,
$\zeta_1=1/2$, $r^2=\Lam^2$ and a modified bare 3-gluon vertex (b)
with $\rho$ decreasing from 1 to 1/2 in the mid-momentum regime in
order to avoid a Landau-type singularity).

Nevertheless, the vertex {\em ansatz}\re{v4A} appears to be too simple
for establishing a full UV-to-IR connection. Although Landau-type
singularities can be avoided over the full momentum window that we
have studied, we have not discovered an IR asymptotics with a high
degree of universality, such as a clear signal of stable power laws for
the dressings.

\subsubsection{IR flow analysis}
\label{IRflow}

Let us now concentrate on the flow of the propagators towards the IR
asymptotics. DSE as well as RG flow equation studies have demonstrated
that the power-law behavior discussed in the introduction is a
self-consistent solution of the dynamical equations in the IR,
implying an IR stable fixed point for the gauge coupling. The
following study is devoted to an investigation of the domain of
attractivity of this fixed point. For this, we assert that the
fluctuations from the UV down to an intermediate scale $\Lam \sim 1
$GeV have already modified the dressing functions as compared to their
perturbative form in an a priori unknown manner.\footnote{In this analysis,
we choose the scale by matching the perturbative one-loop expression for
$\alpha(p^2)$ to our results and employ the experimental input
$\alpha(M_Z)=0.118$ (even though we neglect dynamical quarks). 
The intrinsic uncertainty of such a procedure
is large, but irrelevant for our purposes here.} By assuming various
initial conditions for the dressing functions at intermediate $k$ scales,
we can check whether the flow connects a particular mid-scale initial
condition with the IR 
fixed point regime and power laws. In this way, we can analyze the
mid-momentum requirements that facilitate a full UV -- IR
connection. This particularly provides information about the physical
mechanisms that have to be triggered by the (yet unknown) full
vertices. Throughout this subsection, we use bare ghost and 4-gluon
vertices -- a truncation that is sufficient for the expected IR
asymptotics, owing to ghost dominance in the gluon equation. Beyond
this, we vary the 3-gluon vertex in order to study its influence on
the approach to the IR. 

\begin{figure}
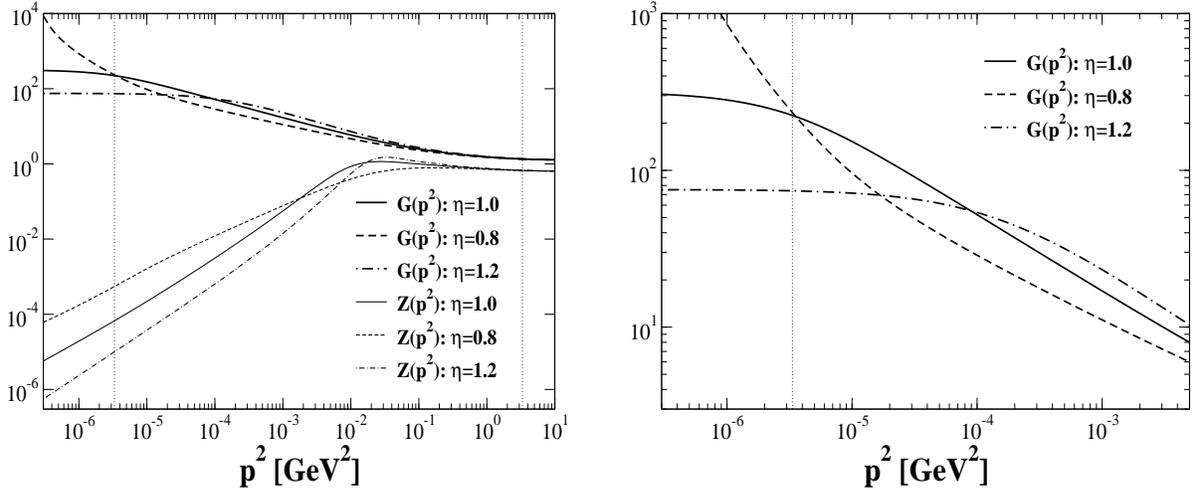

\centerline{
\epsfig{file=igp.dress.ren.eps,width=7.5cm,height=6.5cm}
\hspace{0.5cm} 
\epsfig{file=igp.dress.ren.large.eps,width=7.5cm,height=6.5cm}}

\vspace{-0.3cm} 

\caption{ Numerical results for the ghost and gluon dressing functions
in the minimally dressed vertex truncation (c), Eq.\re{vertex_c},
employing a gluon dressing function with a scale, (\ref{IC3}),
and a constant ghost as
boundary conditions at an intermediate momentum scale. For an initial
gluon power in a small but finite interval around $\eta_A\simeq1$, 
the ghost shows a clear signal for
IR power-law asymptotics with $\kappa\simeq 0.52$. The right panel
magnifies the IR asymptotics of the ghost of the left panel,
demonstrating the failure of the {\em ans\"atze} with $\eta_A=0.8$ or
$\eta_A=1.2$, which are outside the attractive region of the IR-power
law solution. (The vertical dotted line indicates the lower end $k_{\text{IR}}$
of the window of $k$ integration.)}
\label{IRflowfig}
\end{figure}

Two initial conditions for the ghost and gluon dressing functions
at $k=\Lam$ reflecting opposite situations in the mid-momentum regime 
have been given in Eq.\re{IC2} and\re{IC3}. 
In addition to being related to $\Lam$, both boundary conditions contain 
a scale $L$ appearing either in the initial ghost or gluon dressing 
function. Certainly, Yang-Mills theory is governed by only one scale,
which is $\Lambda_{\text{QCD}}$. The relation between $\Lam$ and
$\Lambda_{\text{QCD}}$  is uniquely determined by the RG, once the
coupling $g_\Lam$ is fixed at $\Lam$. Such a unique relation also has
to exist for $L$, i.e., $L=\text{const.} \times
\Lambda_{\text{QCD}}$. To guarantee that our initial condition represents 
a valid approximation to Yang-Mills theory at our initial scale 
$\Lam$, we have to determine this constant. Otherwise our system
describes a different theory with two independent scales. 
    
For solving this problem, we note that, once we have found the right
value of $L$, the flow in the deep IR will not develop yet another
scale and the dressings will depend only on
$p^2/L^2\sim p^2/\Lambda_{\text{QCD}}$. On the other hand, a separate
dependence of our solutions on $L$ and $\Lam$ indicate the failure of
the initial condition. In practice, instead of tuning $L/\Lam$ for fixed
$g_\Lam$, we keep $L/\Lam$ fixed, say $L/\Lam=0.01$, and fine-tune the
value of the initial coupling $g_\Lam$ such that no further scale
arises in our solutions in the deep IR. Examples of such an
additional but unphysical scale are rapid changes, steps or
singularities in the dressings. 

Let us first investigate the initial condition\re{IC2}, which assumes 
that the ghost dressing has already developed a scale, whereas the gluon is 
taken as perturbative 
for all momenta $p^2$. With these boundary conditions we find that 
the flow runs into a singularity at finite $k_{\text{sing}}$ independently 
of the initial ghost exponent $\eta_C$ or the value 
of $L$. The mechanism responsible for such a behavior has been analysed in 
Sect.~\ref{cond-flow}: in the truncation considered here,
the gluon cannot develop an IR suppression with $k \rightarrow 0$.
Thus an unsuppressed gluon at a mid-momentum scale $k=\Lam$
drives the flow away from the IR fixed point.

\begin{figure}
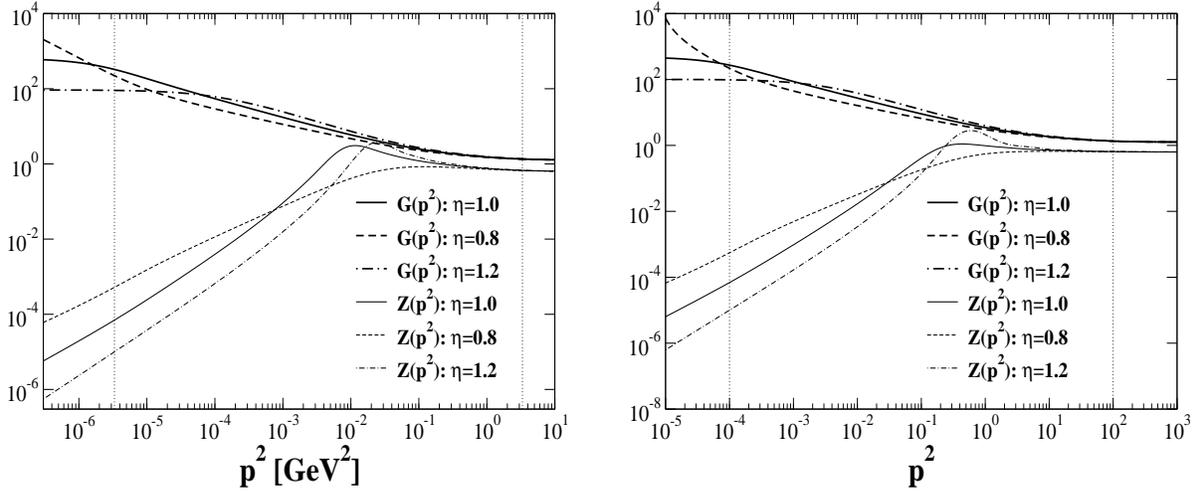

\centerline{
\epsfig{file=igp.ren.eps,width=7.5cm,height=6.5cm}
\hspace{0.5cm} 
\epsfig{file=ghostonly.igp.eps,width=7.5cm,height=6.5cm}}
\caption{ Numerical results for the ghost and gluon dressing
functions, employing a gluon dressing function with a scale, (\ref{IC3}) 
and a constant
ghost as boundary conditions at an intermediate momentum scale. The
left panel shows the result for the modified bare vertex truncation
(b) with $\rho=0.5$, a ``ghost-loop-only'' truncation ($\rho=0$) is
displayed on the right panel. Both truncations lead to IR power law
asymptotics with $\kappa\simeq0.52$ if the initial
gluon power is in a small but finite interval around $\eta_A\simeq1$.}
\label{IRflowfig2}
\end{figure}

This is different for the initial condition\re{IC3}, where we assume 
that the gluon has already developed a suppression with scale $L$
at $k=\Lam$ and the ghost remained perturbative.  
In Fig.~\ref{IRflowfig}, we plot the solution of the flow equation for
the initial conditions with initial gluon exponent $\eta_A=0.8,1.0,1.2$
for the minimally dressed 3-gluon vertex (c), Eq.\re{vertex_c}. The
regulator scale $k$ has been integrated down to a value
$k_{\text{IR}}$ that is indicated by the vertical dotted line on the
left-hand side of the plots. For all three values of $\eta_A$, the
ghost dressings are strongly enhanced in the IR, whereas the gluon
develops a more enhanced peak around $L$.

For $\eta_A=1$ (solid lines), the ghost dressing exhibits a clear
power law, $G(p^2) \sim (p^2)^{-\kappa}$, for all momentum values 
$k_{\text{IR}}^2\lesssim p^2\lesssim L^2$. We identify a ghost exponent of
\begin{equation}
\kappa\simeq 0.52. \label{kappa}
\end{equation}
The gluon develops a power slightly stronger than its initial power
$\eta_A$. This behavior is universal for initial powers $\eta_A$
in a small but finite interval around $\eta_A \simeq 1$, i.e., for
boundary conditions in this interval the IR power $\kappa$ does not 
depend on $\eta_A$. As will be discussed in more detail below, this 
result corresponds to the self-consistent solutions found, for 
instance, in the IR analysis of DSEs.

For $\eta=1.2$ (dot-dashed lines), a power law in the ghost dressing
develops, but not over the full $k$ integration range. At some scale
$k^2\gg k_{\text{IR}}^2$, the ghost gets stuck and becomes essentially
perturbative again. Apparently, the gluon is too strongly suppressed to
control the build up of a power law to arbitrarily small momenta. Since
this solution exhibits a second independent scale where the ghost
becomes perturbative again, we conclude that this {\em ansatz} with
$\eta_A=1.2$ fails to describe the IR of Yang-Mills theory.

For $\eta_A=0.8$, the ghost first builds up a power law, but then
deviates in the IR and runs into a singularity (the plot shows a curve
with $g_\Lam$ chosen such as to allow for a maximal IR extension). For
this set of initial conditions, the gluon is not suppressed strongly
enough in the IR to prevent the same singularity as observed
frequently above. We conclude that also this initial condition with
$\eta_A=0.8$ fails to represent the physics of the IR sector. 

These features remain qualitatively the same for the various 3-gluon
vertices, unless this vertex is not too strong at intermediate
scales. In Fig.~\ref{IRflowfig2}, we show results for the modified
bare vertex (b) with $\rho=0.5$ and a ``ghost-loop-only'' truncation
($\rho=0$). In agreement with the literature, this reveals the ghost
loop as the dominant IR structure, in support of the Kugo-Ojima
confinement scenario. 

\section{Summary and conclusions}

We have performed a study of the vertex expansion of the quantum
effective action of Landau-gauge Yang-Mills theory in the framework of
the exact renormalization group. We have truncated the vertex
expansion at the lowest nontrivial order and concentrated on the
behavior of the ghost and gluon propagators. This level of truncation
already contains important information about large-distance physics 
along the lines of the Kugo-Ojima confinement criterion. 

We identify three different momentum regimes in our investigation: a
perturbative high-momentum region, a nonperturbative regime around
$p^2\sim \mathcal{O}(1)\text{GeV}^2$ presumably dominated by gauge
field fluctuations and a regime in the deep IR dominated by the ghost
degrees of freedom.  Simple truncations of the RG flow equation cannot
only easily deal with the perturbative regime, but give conclusive
answers for the deep IR. In the literature, a power-law behavior of
the dressings has been self-consistently determined in the deep IR,
which is related to an IR fixed point of the running coupling and
realizes the Kugo-Ojima confinement criterion
\cite{Lerche:2002ep,Zwanziger:2001kw,Fischer:2002hn,Pawlowski:2003hq}.

In addition to these studies, our work analyzes the flow towards this
fixed point and explores its domain of attractivity. For this, we
employ a set of mid-scale initial conditions for the propagator
dressings that may arise from the flow, once the fluctuations at
intermediate momenta have also been properly integrated out. Within this
set, our results demonstrate that a mass-like structure in the gluon
dressing (characterized by an initial gluon power in a small but 
finite interval around $\eta_A\simeq1$ and a scale $L$) is a
necessary prerequisite for approaching the power-law asymptotics in
the IR. Other tested mid-scale initial conditions lead to either a
singular or a trivial behavior of the propagator dressings. For the 
mass-like
structure, the resulting flow shows a universal behavior, being
largely independent of the initial details, such as further properties
of the initial conditions or the form of the gluonic vertices.

As a result, we observe a ghost exponent of $\kappa\simeq0.52$. Even
though the corresponding value of the DSE IR analysis is
$\kappa\simeq0.595$, these results are in satisfactory agreement,
since the exponents are regulator dependent in the present
truncation. This regulator dependence can quantitatively be studied in
a self-consistent IR analysis in the flow equation framework
\cite{Pawlowski:2003hq}, revealing $\kappa\simeq0.539$ for the
regulator used in the present work \cite{PawlowskiPrivate}; the
difference to our result obtained by solving the flow equation gives a
measure for the numerical accuracy of our procedure.\footnote{In fact,
since the flow has to build up the full ghost power from an initially
flat dressing, numerical errors generically accumulate such as to
reduce the final value of $\kappa$.}

We furthermore have searched
for global solutions connecting both asymptotic ends of the momentum
range. Here the mid-momentum regime appears most difficult and
represents a delicate problem. We show that oversimplified truncations
based on bare vertices lead to Landau-type singularities in the flow,
signaling the importance of higher-order correlations. Dynamical
information encoded in dressed (or ``running'') vertices is required
in this mid-momentum regime.

In the present work, we have modeled this dynamical information by
supplementing the bare-vertex index structures with (positive)
momentum-dependent dressing functions. Within these limits, we observe
that a truncation on the 3-point level typically leads to ghost {\em
and} gluon enhancement in the IR, and is therefore insufficient to
describe the IR suppression of the gluon dressing function, as, e.g.,
observed on the lattice
\cite{Langfeld:2001cz,Bonnet:2001uh,Bowman:2004jm}. Once suitable
momentum-dependent 4-point dressings are taken into account, the gluon
does indeed exhibit the expected IR suppression. We have demonstrated
this property by garnishing the 4-gluon vertex with generic powers of
the propagator dressing functions. However, even though the resulting
flow remains stable over all momentum scales of the numerical
integration, the resulting propagators in the deep IR depend strongly
on the details of the modeling with no sign of the expected universal
behavior. We conclude that in the truncations considered, important
dynamical information from the mid-momentum regime is still missing.

As a main conclusion, we interpret our results as follows: the
IR power-law asymptotics of the propagators is triggered (though not
generated!) by a particular dynamics at intermediate
momenta. If this dynamics gives rise to a mass-like structure in the
gluon propagator, the Kugo-Ojima branch acts as an IR attractive fixed
point of the flow, resulting in ghost enhancement and gluon
suppression in a universal manner. Our results illuminate the role
played by the mid-momentum regime, but, apart from some
qualitative insight, the physical mechanisms potentially governed by
higher-order correlations are still poorly understood. 

Concerning quantitative estimates made in this paper, let us note that
we have expressed all momenta in physical units (GeV) by fixing the
flow at the $Z$ mass, $M_Z$. We have done so mainly for illustrative
purposes. We stress that resulting scales in the nonperturbative
sector can sizably be modified by changes in the vertex dressings and, 
of course, upon the inclusion of dynamical quarks. 

From a conceptual perspective, an important ingredient for our
findings is the modified Ward-Takahashi identity (mWTI). Beyond the
dynamical information contained in the flow equation, the mWTI acts as
an additional constraint that encodes gauge invariance even in the
presence of a gauge-symmetry breaking regulator. In the present case,
we have employed the mWTI for controlling the running of the gluon
mass $m_k$ at finite $k$; this turns the gluon mass from a seemingly 
independent and relevant RG parameter into a dependent and RG irrelevant 
quantity. In the language of loop integrations, the gluon mass  
corresponds to quadratic divergencies arising in non-gauge-invariant 
regularizations for which appropriate subtraction procedures have to be 
defined. We have verified that naive subtraction procedures not obeying 
the mWTI generically generate an artificial mass scale in the flow. This in 
turn also leads to a global solution which smoothly interconnects the 
perturbative and the IR power-law behavior. However it does so for the 
wrong mechanism. The scale $L$ discussed above certainly should be 
generated dynamically and not by the regularization procedure.

Future work should be devoted to an understanding of the dynamical
mechanisms that initiate gluon suppression, as required for
approaching the IR attractive domain of the Kugo-Ojima branch. 
A possible route is the inclusion of the next order in the vertex
expansion, pursuing a dynamical calculation of vertices. Promising
work in this direction in the flow equation framework has been done in
\cite{Kato:2004ry}, where an attempt was made to compute
momentum-dependent 4-point vertices. The resulting quantitative
deviations from other results in the literature may be associated
with the use of an angular approximation and the neglect of the
3-gluon vertex, but improvements are certainly straightforwardly
amenable. 

Finally, it remains to be understood how the vertex expansion can be
connected with another nonperturbative truncation scheme: the
expansion of the effective action in field-strength
invariants \cite{Reuter}. Whereas the vertex expansion concentrates on
the flow of operators with a small number of fields but full momentum
dependencies, the expansion in terms of field-strength invariants can
deal with an infinite number of fields but usually neglects momentum
dependencies. It is reassuring that an IR fixed point in the coupling
has also been observed in the latter expansion of the effective action
\cite{Gies:2002af}, in particular, since both approaches exploit a
non-renormalization theorem in order to define the running coupling.

\section*{Acknowledgment}

We are grateful to J.M.~Pawlowski for a series of intense discussions 
and for providing numerical information about the regulator dependence 
of $\kappa$. We would like to thank R.~Alkofer, J.~J\"ackel and 
J.M.~Pawlowski for a critical reading of the manuscript. We are  
grateful to R.~Alkofer, J.~Berges, U.~Ellwanger and J.~J\"ackel for 
helpful discussions.  This work was supported by the Deutsche 
Forschungsgemeinschaft (DFG) under contracts Gi 328/1-2 (Emmy-Noether
program) and Fi 970/2-1. 

\begin{appendix}

\renewcommand{\thesection}{\mbox{\Alph{section}}}
\renewcommand{\theequation}{\mbox{\Alph{section}.\arabic{equation}}}
\setcounter{section}{0}
\setcounter{equation}{0}

\section{Integral kernels}
\label{kernels}
Here we give the explicit formulae for the momentum integral kernels
of the propagator equations\re{fm1} and\re{fm2}.  In the Landau gauge,
only the kernels involving transversal gluons are of primary importance. For
the bare-vertex truncation, they read
{\small
\begin{eqnarray}
Q^{\ZT,\ZT}(\xh,\yh,u)&=& -(1-u^2) \left[ x+y+
\frac{xy}{z} \left( 1-\frac{1-u^2}{D-1} \right) \right],
\label{A.1}\\
Q^{G,G}(\xh,\yh,u)&=&\frac{\yh(1-u^2)}{2(D-1)},\label{A.2}\\
Q^{G,\ZT}(\xh,\yh,u)&=& \frac{\left(1- u^2 
\right) \,\xh\,\yh}{\zh},\label{A.3}
\end{eqnarray}
}
where the variable $\zh$ on the right-hand side has to be understood as
an abbreviation of $\zh=\xh+\yh-2u\sqrt{\xh\yh}$.

\section{Regulator properties}
\label{cutoffs}
\setcounter{equation}{0}

For the regulator shape function specified in Eq.\re{2.8},
\begin{equation}
r(\xh)\equiv r_{A,C}(\xh)=\frac{1}{\xh} (1-\xh)\, \theta(1-\xh),\label{B1}
\end{equation}
the following identities hold:
\begin{equation}
\xh r(\xh) =(1-\xh)\, \theta(1-\xh), \quad -\xh^2
r'(\xh)=\theta(1-\xh). \label{B2}
\end{equation}
These are required in Eq.\re{Kreg}. The regularized propagators
finally read
\begin{eqnarray}
\frac{1}{p_A(\xh)}&=&\frac{\theta(1-\xh)}{1+\ZT(\xh) \: \mqk/k^2}
	 +\frac{\theta(\xh-1)}{\xh+ \ZT(\xh) \: \mqk/k^2}, \quad
	 \ \nonumber\\
\frac{1}{p_C(\xh)}&=&\theta(1-\xh)
	 +\frac{\theta(\xh-1)}{\xh}.\label{B3}
\end{eqnarray}

\section{RG rescaling}
\label{scaling}
\setcounter{equation}{0}

Let us study the system under (finite) RG rescalings of the renormalized 
fields of the form
\begin{equation}
A_\mu \to \za^{-1/2}\, A_\mu,\quad (C,\cbar)\to \zc^{-1/2}\,
(C,\cbar), \label{sc1}
\end{equation}
where $\za,\zc$ account for a possibly $k$-dependent RG
scaling. Invariance of the action under this scaling transformation
implies that the $n$-point vertices scale as
\begin{equation}
\Gamma^{(n+m)}_{A^n,C^m} \to \za^{n/2} \zc^{m/2}\,
\Gamma^{(n+m)}_{A^n,C^m}. \label{sc2}
\end{equation}
The flow equation is invariant for regulators that scale as
\begin{equation}
R_A \to \za\, R_A, \quad R_C\to \zc\, R_c. 
\label{sc3}
\end{equation}
As a special case of Eq.\re{sc2}, let us note that the propagator
dressing functions scale as
\begin{equation}
G(x) \to \zc^{-1}\, G(x), \quad \ZT(x)\to \za^{-1}\,
\ZT(x). \label{sc4}
\end{equation}
Owing to its nonperturbative definition\re{coupling}, the renormalized
gauge coupling scales as
\begin{equation}
g^2\to \zc^2\za\, g^2, \label{sc5}
\end{equation}
such that $g^2 G^2\ZT$ is invariant as required. In Eq.\re{2.5}, we
defined the vertex dressings $V$ by scaling out the coupling,
$\Gamma^{(3)}=g\, V^{(3)}$, $\Gamma^{(4)}=g^2\, V^{(4)}$; hence the
vertex dressings have to scale as
\begin{equation}
V_{\cbar AC}\to V_{\cbar AC},\quad V_{3A}\to \za \zc^{-1}\,
V_{3A},\quad V_{4A}\to \za\zc^{-2}\, V_{4A}.\label{sc6}
\end{equation}
These properties tell us, for instance, that the naive bare vertex
truncation $V_{3A}\to 1$ is generally not compatible with RG
scaling, and therefore only correct if the fields are normalized
canonically. 

\section{Numerical procedure \label{numproc}}

For the numerical computations, we have to specify the regulator shape
function introduced in Eq.\re{2.7}. Here we use the one proposed in
\cite{Litim:2001up},  
\begin{equation}
r_{A,C}(\xh)=\frac{1}{\xh} (1-\xh)\, \theta(1-\xh).\label{2.8}
\end{equation}
The step function in Eq.\re{2.8} leads to a more localized support of
the loop momentum integrations and thereby simplifies the required
algorithms.\footnote{Whereas the nonanalyticity of the $\theta$
function can become problematic in high-order derivative expansions
\cite{Canet:2002gs}, there is no conflict with the vertex expansion
and the properties of this shape function will remain numerically
useful also to higher orders.}

For the actual numerical integration of the flow from the starting scale 
$\Lam$ to an IR scale $k_{\text{IR}}$, we employ a fifth-order 
Runge-Kutta formula with an embedded fourth-order formula to estimate 
integration errors. We have also tested Euler's method and found it
neither accurate nor stable enough for the present problem. The loop 
integrals on the right-hand side of the flow equations are carried out 
numerically; no angular approximations are made. To this end, the 
dressing functions have to be represented over the whole momentum range. 
We choose a Chebyshev expansion on a logarithmic momentum grid, 
$p^2 \in [\epsilon^2_p,\Lambda^2_p]$ with $\epsilon^2_p \ll k^2_{IR}$ and 
$\Lambda^2_p \gg \Lam$. An expansion of the dressing functions in terms 
of the first $N$ Chebyshev polynomials is exact for those $N$ momenta 
that are zeros of the $(N+1)$st polynomial. Thus, taking these zeros as 
external momenta $p^2$ in the RG equations ensures maximal numerical 
stability. A typically chosen value for $N$ is $N=45$. For momenta far below 
$\epsilon^2_p$, we extrapolate the Chebyshev expansion employing either
a power-law fit or a smooth continuation to a constant value; the
latter is justified by the presence of the regulator. We have
confirmed that our results do not depend on the details of 
this extrapolation. 
In general, the advantage of the Chebyshev representation over 
the parametrization method used in \cite{Ellwanger:1996wy} is that one 
does not rely on a specific form of fit functions in the interval 
$[\epsilon^2_p,\Lambda^2_p]$ which could possibly bias the results. 

\end{appendix}
\goodbreak

\end{document}